\documentstyle[12pt,epsfig]{article}

\newcommand{\R}[1]{{\mathrm #1}}

\newlength{\extraspace}
\newlength{\extraspaces}
\newcommand{\beq}{\begin{equation}}
\newcommand{\eeq}{\end{equation}}
\newcommand{\be}{\begin{equation}}
\newcommand{\ee}{\end{equation}}
\newcommand{\bea}{\begin{eqnarray}}
\newcommand{\nn}{\nonumber}
\newcommand{\eea}{\end{eqnarray}}
\newcommand{\rf}[1]{(\ref{#1})} 

\newcommand{\ra}{\rightarrow}

\newcommand{\half}{{1\over 2}}
\newcommand{\smallhalf}{{\textstyle{1\over 2}}}
\newcommand{\smallthreehalves}{{\textstyle{3\over 2}}}

\newcommand{\fivehalves}{{5\over 2}}
\newcommand{\fivequarters}{{5\over 4}}

\newcommand{\quarter}{{1\over 4}}

\newcommand{\expect}[1]{{\langle #1\rangle}}

\newcommand{\zbar}{{\bar z}}
\newcommand{\ket}[1]{{\vert#1\rangle}}
\newcommand{\bra}[1]{{\langle#1\vert}}

\renewcommand{\Im}{{\R {Im}}\,}
\renewcommand{\Re}{{\R {Re}}\,}

\begin{document}

\thispagestyle{empty}

\begin{flushright}
{\sc OUTP}-00-05P\\
21st  March 2000\\
revised 24th June 2000\\
 hep-th/0003184
\end{flushright}
\vspace{.3cm}

\begin{center}
{\large\sc{Boundary Logarithmic Conformal Field Theory} }\\[15mm]

{\sc Ian  I. Kogan\footnote{e-mail: i.kogan1@physics.ox.ac.uk}

 and John  F. Wheater\footnote{e-mail: j.wheater1@physics.ox.ac.uk}} \\[5mm]
{\it Department of Physics, University of Oxford,\\Theoretical Physics,\\ 1 
Keble Road,\\
       Oxford, OX1 3NP, UK} \\[15mm]

{\sc Abstract}

\begin{center}
\begin{minipage}{14cm}
We discuss the effect of boundaries in boundary 
 logarithmic conformal field theory and show, with reference
to both  $c=-2$ and $c=0$ models, how they produce new features
even in bulk correlation functions which are not present in the
corresponding models without boundaries. We discuss the modification
of Cardy's
relation between boundary states and bulk quantities.
\end{minipage}
\end{center}

\end{center}

\noindent

\vfill\noindent
PACS codes: 11.25.Hf, 11.10.Kk\\
Keywords: logarithmic conformal field theory

\newpage
\pagestyle{plain}
\setcounter{page}{1}

\renewcommand{\footnotesize}{\small}
\section{Introduction}

Conformal field theories (CFT) are of great importance in  modern theoretical 
physics. Some of the most spectacular progress in the last 15 years has
been  in our understanding of two-dimensional conformal field theories
 which play an important role in string theory, statistical mechanics
and condensed matter physics. Immediately after the first paper by
Belavin, Polyakov and Zamolodchikov \cite{BPZ} in which  it was shown
 how conformal invariance in two dimensions can completely determine
 the critical exponents and  bulk correlation functions,  Cardy
\cite{cardy1}  showed how conformal symmetry can determine critical 
exponents and correlation functions in the presence of a boundary.
Boundary conformal field theories can be defined in any number of
dimensions $d$  and one can get some general results  for any $d$,
but the strongest results, of course, are found for $d=2$.
The main result of \cite{cardy1} was that the $n$-point
correlation function in the presence of a boundary   satisfies
 the same equation as the $2n$-point correlation function in the bulk,
provided one chooses conformal boundary conditions.  Subsequently
it was  understood  how to classify  different
boundary conditions and how to relate bulk and boundary operators 
\cite{cardy2,ishibashi}.
Boundary CFT is of interest not only to the condensed matter
community where systems with boundaries are obviously important
but also 
 for the string community, because it gives a mathematical framework
to formulate the theory of open strings  \cite{openstrings,ps}
(and more recently D-branes \cite{Dbrane}).
More complete references are given in \cite{tensionasdimension}.

More recently Gurarie \cite{gurarie} 
drew attention to  logarithmic
conformal field theories (LCFT).  
 In LCFT there are logarithmic terms in some 
correlation functions but the theories are nonetheless 
 compatible with conformal invariance. 
An LCFT appears when two (or more, but this is not the general case)
operators become degenerate and form a logarithmic pair, 
usually denoted $C$ and $D$.
 The OPE of the stress-energy tensor T with the
   logarithmic operators $C$ and $D$ 
 is non-trivial and involves mixing~\cite{gurarie}
\bea
&~&T(z) C(w) \sim  \frac{h}{(z-w)^2}C(w) +
 \frac{1}{(z-w)} \partial_z C \dots \nn \\
&~&T(z)D(w) \sim \frac{h}{(z-w)^2} D(w) + \frac{1}{(z-w)^2}C(w) + 
   \frac{1}{(z-w)} \partial_z  D +
\dots
\label{stress}
\eea
where $h $ is the conformal dimension of the
operators with respect to the holomorphic stress-energy tensor $T(z)$.
The OPE with $\bar{T}$ has the same form  but with $\bar{h}$ instead of
$h$; as usual the scaling dimension is $h+\bar{h}$ and the
 spin of the field is $h-\bar{h}$.

It is a consequence of \rf{stress}
that under a conformal transformation $ z \ra  w= z +
\epsilon(z)$ the  logarithmic pair is  transformed  as
\bea
\delta C =\partial_z\epsilon(z)h C + \epsilon(z) \partial_z C + \dots
\nn \\
\delta D =\partial_z\epsilon(z)(h D +  C) + \epsilon(z)
\partial_z D + \dots
\eea
which  can be written globally as 
\bea
\left(\begin{array}{c}
C(z) \\ D(z)
\end{array}\right)
= \left(\frac{\partial w}{\partial z}\right)^{\ \left(\begin{array}{cc}
h & 0 \\
 1 & h
\end{array}\right) }
\left(\frac{\partial \bar{w}}{\partial \bar{z}}\right)^{\left(\begin{array}{cc}
\bar{h} & 0 \\
 1 & \bar{h}
\end{array}\right) }
\left(\begin{array}{c}
C(w) \\ D(w)
\end{array}\right)
\eea
From this conformal transformation one can derive 
the two point functions  for the logarithmic pair  \cite{gurarie,ckt}
\begin{eqnarray}
\langle C(x) D(y)\rangle &&= 
\langle C(y) D(x) \rangle  = \frac{\alpha_D}{(x-y)^{2h }}\nonumber \\
\langle D(x) D(y)\rangle &&= 
 \frac{1}{(x-y)^{2h}} \left(-2\alpha_D\ln(x-y) + \alpha_D'\right)
\nonumber \\ 
\langle C(x) C(y)\rangle  &&= 0
\label{CC}
\end{eqnarray}
 where the constant $\alpha_D$ is determined by the normalization of the $D$
operator and the constant $\alpha_D'$ can be changed by the redefinition
 $D \ra D + C $. Note that (\ref{CC}) is absolutely universal and valid in any
number of dimensions, because only the most general properties of
conformal symmetry were used to derive it. One can easily generalize
these formulas to the case when there are $n$ degenerate
fields and the Jordan cell is given by an $n \times n$ matrix, in which
case the maximal power of the logarithm will be $\ln^{n-1} z$; some
explicit expressions can be found, for example,  in \cite{higherdimensions}.

Much is known about the general
properties of these theories; for example, the presence of
a zero norm state \cite{ckt}, the fusion rules and 
 modular properties \cite{fusionrules1,fusionrules2,fusionrules3}, the Couloumb gas description
 of LCFT \cite{c=-2,otheralgebras}, the existence of logarithmic
pairs with respect to other algebras such as affine Lie algebras
 \cite{otheralgebras,gravgaugedressing}, and the  emergence of
 LCFT in $c=0$ theories in general \cite{c=0 Quenched Random Magnets
and Polymers}.
LCFTs have applications in many  areas; for example, percolation
\cite{percolation}, the WZNW model on the supergroup $GL(1,1)$ \cite{rs},
 gauge and gravitational  dressings of
non-logarithmic CFT \cite{gravgaugedressing}, 
the world-sheet description of soliton collective coordinates in
string theory and D-brane recoil \cite{strings},  disordered conductors
and the Quantum Hall Effect\cite{disorder}, planar magnetohydrodynamics
\cite{hydrodynamics}, and some supersymmetric WZNW models
\cite{supersymmetric}.  Their  deformation by marginal and slightly 
relevant logarithmic operators was studied in \cite{ckt,RG}.
There are several interesting ``holographic'' relations between
$d$-dimensional LCFT  on a boundary and $d+1$ dimensional bulk
theories \cite{holographic} as well as with Seiberg-Witten theory
\cite{seibergwitten}.

Most of the literature  is concerned with the bulk properites
 of LCFT.
However, boundary problems appear in a number of important
applications; in the D-brane recoil
problem \cite{strings} the   recoil  operators  must
be boundary logarithmic operators and it is natural to consider
percolation and disordered systems 
in the presence of boundaries \cite{c=0 Quenched Random Magnets and Polymers}.
In this letter we discuss several basic properties of boundary
LCFT (see also \cite{snooks}), and  how the  methods of  ordinary boundary CFT can
be generalised to the LCFT case.

\section{Two-point correlation functions in the presence of boundary}
Let us consider CFT on the upper half-plane $\Im z \geq 0$ (Fig.1).
As was shown in  \cite{cardy1},
 two-point functions in the presence of the  boundary
  are
related to  four-point functions on the whole plane provided 
the boundary conditions are conformally invariant 
so that $T = \bar{T}$ when $\Im z =0$.
 These  boundary 
conditions  allow us to analytically continue $T$ from the
upper half plane to the whole plane by setting
 $T(z)$ for $\Im z <0$  to $\bar{T}(\zbar)$.
One can then show that by combining two contours $C$ and $\bar{C}$ (see 
Fig.1) into one
 on the whole plane that  the $n$-point function in the presence
of the  boundary
 $\expect{\Phi(z_1, \bar{z}_1) \Phi(z_2, \bar{z}_2) \dots
 \Phi(z_n, \bar{z}_n)}$  which is  a function of $2n$ variables
 $(z_1,z_2,...z_n,\bar{z}_1,...\bar{z}_n)$ satisfies the same
 differential equation as the $2n$-point functions of the same CFT on the 
 whole plane  $\expect{\Phi(z_1, \bar{z}_1) \Phi(z_2, \bar{z}_2) \dots
 \Phi(z_{2n}, \bar{z}_{2n})}$, regarded as a function of holomorphic variables
 $(z_1,.... z_{2n})$ only.

\begin{figure}[ht]
\begin{center}
\mbox{\epsfig{file=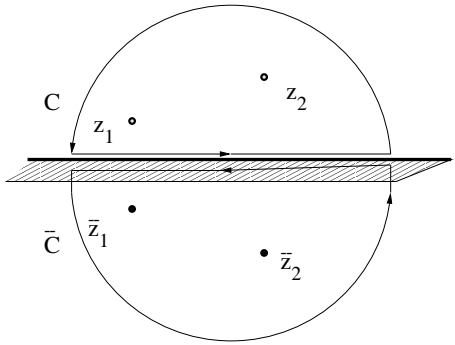,width=6cm}}
\caption { Contours $C$ and $\bar{C}$ together make a  contour
 on the full plane encircling all four points $z_1, z_2, z_3 = \bar{z}_2$ and
$z_4 = \bar{z}_1$ thus establishing the  relation between 2-point functions
on the  half plane and  4-point functions on the whole plane.}
\label{Fig.1}
\end{center}
\end{figure}

Specializing to two-point functions we see immediately that they
are not yet completely determined by this construction because there
are two solutions to the differential equation for the four
point function; the correct combination will be determined by the
boundary conditions.
This immediately leads to a very interesting fact.   For fields
which give logarithmic operators as the result of fusion, for example
\bea
\mu \times \mu = C + D,
\eea
 logarithmic correlations can be observed only  for four-point and
higher order correlation functions in bulk LCFT. However when a 
boundary is present we can get logarithmic terms  in 
the {\it two-point function}  because $\expect{\mu \mu}$ is related to the bulk
 four-point
function $\expect{\mu \mu \mu \mu}$;  the very existence of the  boundary  leads
to this new behaviour.

To study this in more detail  consider the $c=-2$ theory first.
In the bulk the chiral part of the
 four-point function for the $(1,2)$ operator $\mu(z,\zbar)$ with
dimension $-1/8$  can  be defined from Ward identities with respect to 
$T$ and is given by
\bea \lefteqn{\expect{\mu(z_1,\zbar_1)\mu(z_2,\zbar_2)\mu(z_3,\zbar_3)\mu(z_4,\zbar_4)
}_{chiral}=}\nn \\
& &(z_1-z_3)^\quarter(z_2-z_4)^\quarter
 (\xi(1-\xi))^\quarter 
\left(A F(\smallhalf,\smallhalf;1;\xi)+B F(\smallhalf,\smallhalf;1;1-\xi)\right)\label{4chiral}\eea
where we have chosen the anharmonic ratio
\beq \xi=\frac{z_{12}z_{34}}{z_{13}z_{24}}.\eeq
The constants $A$ and $B$ depend on $\bar{z}_1,..\bar{z}_4$ and using the $\bar{T}$
 Ward identities we  see that there must be the same functional 
dependence on $\bar{z}$, i.e. $A$ and $B$ must be  superpositions
of $ F(\half,\half;1;\bar{\xi})$ and $ F(\half,\half;1;1-\bar{\xi})$.
 Because the full left-right symmetric correlation function must  be
free of logarithmic cuts there is no ambiguity in constructing the
full answer (see Saleur \cite{percolation})
\bea
\lefteqn{\expect{\mu(z_1,\zbar_1)\mu(z_2,\zbar_2)\mu(z_3,\zbar_3)\mu(z_4,\zbar_4)
}= 
|z_1-z_4|^{1/2}|z_3-z_2|^{1/2}
 |\xi(1-\xi)|^{1/2}} \nn \\
&&\times\left( F(\smallhalf,\smallhalf;1;\xi) F(\smallhalf,\smallhalf;1;1-\bar{\xi})
+ F(\smallhalf,\smallhalf;1;1-\xi)F(\smallhalf,\smallhalf;1;\bar{\xi})\right).
 \label{4full}\eea

 Now  consider the two-point function for the  same $(1,2)$ operator
 in the presence
of a boundary along the real axis. As discussed above
  it is given by the solution to the differential 
equation for the  holomorphic part of the four point function 
without a boundary  (\ref{4chiral}). We identify $z_3$ with $\zbar_2$
and $z_4$ with $\zbar_1$ so that 
\beq \xi=\frac{\vert z_1-z_2\vert^2}{\vert z_1-\zbar_2\vert^2}\eeq
and is always between 0 and 1. Then the two point function is given by
\bea\lefteqn{ \expect{\mu(z_1,\zbar_1)\mu(z_2,\zbar_2)}_{boundary} =}\nn\\
& &(z_1-\zbar_2)^\quarter(\zbar_1-z_2)^\quarter
 (\xi(1-\xi))^\quarter\left(A F(\smallhalf,\smallhalf;1;\xi)+B
 F(\smallhalf,\smallhalf;1;1-\xi)\right)\label{boundary2ptfn}\eea
and since  the hypergeometric function has a cut along $[1,\infty]$
this expression is always well-defined and real in the physical
region.
If we let the points $z_1$ and $z_2$ move away from the boundary but keep
their separation fixed then $\xi\to 0_+$  and we see that the first term in
the solution gives a contribution which is like the bulk two-point
function
\beq \expect{\mu(z_1,\zbar_1)\mu(z_2,\zbar_2)}=Az_{12}^\quarter \zbar_{12}^\quarter.\eeq
On the other hand the second term contains a  logarithmic piece 
\beq \expect{\mu(z_1,\zbar_1)\mu(z_2,\zbar_2)}=Bz_{12}^\quarter \zbar_{12}^\quarter\log\vert z_{12}\vert^2.\eeq
 One might argue that in order to recover the standard bulk two-point 
function, which does not contain a logarithm, when 
the points are far from the boundary we should  set $B=0$. 
In a unitary theory this would be a possible solution
 but here it is not at all clear because 
the theory is non-unitary 
and the bulk two-point function grows with separation.
  Thus there is no physical motivation for supposing that when 
$z_1$ and $z_2$ are far from the boundary
the correlation function is unaffected by the operators
at the image points -- in general it clearly is. 
At the other extreme we let the  points $z_1$ and $z_2$ approach
 the boundary so that $\Im z_1=\Im z_2=y\to 0$ while keeping their separation $x$ fixed;  we now have $\xi=(1+\frac{4y^2}{x^2})^{-1}$ approaching
1.  Now the second term in the two point function \rf{boundary2ptfn}
 displays regular
power law behaviour while the first term, which is regular in the bulk,
gives the logarithmic behaviour
\beq \expect{\mu(z_1,\zbar_1)\mu(z_2,\zbar_2)}=A(4y^2)^\quarter
\log\frac{4y^2}{x^2}.\eeq
The constants $A$ and $B$ in \rf{boundary2ptfn} 
must be determined by the boundary
conditions; however, we see that whatever these are,
 logarithmic terms must appear either in the bulk 
or near the boundary.

  This phenomenon is not confined
to the $c=-2$ model. It appears also in the $c=0$ model
describing the  percolation problem considered by Gurarie and Ludwig
\cite{c=0 Quenched Random Magnets and Polymers}.
For example, the two point function
of the bulk energy operator $\epsilon(z,\zbar)$ which has
conformal dimension $5/8$ is given 
in the  upper half plane by
\beq \expect{\epsilon(z_1,\zbar_1)\epsilon(z_2,\zbar_2)}=
\frac{1}{\vert z_1-z_2\vert^\fivehalves(1-\xi)^\fivequarters}
\left(B(1-\xi)^2 F(-\smallhalf,\smallthreehalves;3;1-\xi)+A\xi^2
 F(-\smallhalf,\smallthreehalves;3;\xi)\right).\eeq
When the operators are far from the boundary and $\xi$ is small, the 
first term gives logarithmic 
 behaviour 
\beq \expect{\epsilon(z_1,\zbar_1)\epsilon(z_2,\zbar_2)}=
 \vert z_{12}\vert^{-\fivehalves}\left(1+\frac{15}{32}\xi^2\log \xi+\ldots\right).\eeq
 This logarithmic
behaviour is what is expected
for the bulk two point function which in this case declines with distance
so we are justified in ignoring the effect of the boundary and concluding that
$B=1$. On the other hand when the operators are close to
 the boundary and $\xi$ approaches 1 we see that the second term, whose coefficient $A$ is not fixed by considering the bulk correlation function,
 gives logarithmic behaviour.

Another interesting example 
at  $c=0$   is the $k=0$  $SU(2)$ WZNW model which is the bosonic sector of the
$N= 1$  SUSY $SU(2)$ WZNW model at $k=2$. This theory is logarithmic
\footnote{The general  case of $SU(N)$ at level $k=0$ was discussed in  KM 
\cite{strings}  and the $SU(2)$ case was discussed in more detail in
CKLT  \cite{supersymmetric}. }
but contains no negative dimension operators.
 The chiral four-point function is given by
\bea
\lefteqn{\expect{ V_{\epsilon_1} (z_1,\zbar_1) V_{\epsilon_2}
(z_2,\zbar_2) V_{\epsilon_3} (z_3,\zbar_3) V_{\epsilon_4}(z_4,\zbar_4)}_{chiral}=} \nn \\
 & &(z_1-z_3)^{-3/4}(z_2-z_4)^{-3/4}
 (\xi(1-\xi))^{1/4} \left(A\sum_{i=1}^{2} J_{i} F^{i}_{A}(\xi)
 + B\sum_{i=1}^{2} J_{i} F^{i}_{B}(\xi)\right)
\label{su20chiral}\eea
where $V_{\epsilon}$ is a primary chiral field in the 
fundamental representation, $\epsilon = \pm 1$,
$ J_1 = \delta_{\epsilon_1 \epsilon_2}\delta_{\epsilon_3 \epsilon_4},
 J_2 = \delta_{\epsilon_1 \epsilon_4}\delta_{\epsilon_2 \epsilon_3}$ and
  $\sum_{I=1}^{4}\epsilon_I = 0$.
 The functions $F^{I}_{A,B}(\xi)$ are given by
\begin{eqnarray} 
F_A^{1}(z) &=& F(1/2, 3/2; 1; \xi) \nonumber \\
F_B^{1}(z) &=&F(1/2, 3/2; 2; 1-\xi) \nonumber \\
&=&-\frac{2}{\pi} \ln \xi F(1/2, 3/2; 1; \xi) - \frac{2}{\pi} H_{0} (\xi)  \nonumber \\ 
F_A^{2}(\xi) &=&F(1/2, 3/2; 2; \xi) \nonumber \\
F_B^{2}(\xi) &=& 2 F(1/2, 3/2; 1; 1-\xi) \nonumber \\
&=& \frac{4}{\pi\xi} -\frac{1}{\pi} \ln \xi F(1/2, 3/2; 2; \xi) -\frac{1}{\pi}
H_{1} (\xi)  \nonumber \\ 
H_{i} (\xi) &=& \sum_{n=0}^{\infty} \xi^n \frac{(1/2 )_n (3/2)_n}{n! (n+i)!}
\times  \left\{ \Psi (1/2 + n)  + \right. \nonumber \\ 
  && \left. + \Psi (3/2 + n) - \Psi (n+1) - \Psi (n+i+1) \right\} 
\end{eqnarray}
 The functions $F_A^{i}$ and $F_B^I$ have logarithmic behavior near 
$\xi = 1$ and  $\xi = 0$ respectively
It is easy to see that one must have the following OPE
\bea
V_{\epsilon_1} (z_1) V_{\epsilon_2} (z_2) = \frac{1}{z_{12}^{3/4}}
\left\{ I_{\epsilon_1 \epsilon_2^{\vee}} - 
z_{12} t^i_{\epsilon_1 \epsilon_2^{\vee}} \left[ D^i
(z_2) +
\ln z_{12} C^i (z_2) \right] + ... \right\} \label{OPEVV}
\eea
where $I$ is the unit matrix and $\epsilon^{\vee}$ is the weight 
conjugate to $\epsilon$. We see that logarithmic operators are transformed as
 a conjugate representation and have dimension $ 2/(k+2) = 1$. 
We can now write the  two-point functions 
\bea
\expect{ V_{+} (z_1,\zbar_1) V_{+} (z_2,\zbar_2)}_{boundary}= 
\expect{ V_{-} (z_1,\zbar_1) V_{-} (z_2,\zbar_2)}_{boundary}=
\nn \\
(z_1-\zbar_2)^{-3/4}(\zbar_1-z_2)^{-3/4}
 (\xi(1-\xi))^{1/4} 
 \left(A F(1/2, 3/2; 1; \xi)
 + B F(1/2, 3/2; 2; 1-\xi) \right)
\label{su20boundary++}
\eea
 and
\bea
\expect{ V_{+} (z_1,\zbar_1) V_{-} (z_2,\zbar_2)}_{boundary}= 
\expect{ V_{-} (z_1,\zbar_1) V_{+} (z_2,\zbar_2)}_{boundary}= 
\nn \\
(z_1-\zbar_2)^{-3/4}(\zbar_1-z_2)^{-3/4}
 (\xi(1-\xi))^{1/4} \left(\frac{A}{2} F(1/2, 3/2; 2; \xi)
 + 2 B F(1/2, 3/2; 1; 1-\xi) \right)
\label{su20boundary+-}
\eea
Again, the same general features emerge. Whatever the boundary
conditions at the very least there will be logarithmic behaviour 
either in the bulk or near the boundary, if not both.

\section{Bulk and boundary operators in LCFT}
When we compute a correlation function in the boundary theory
for every   bulk operator $\Phi(z_1)$ on the upper half plane
 there is a   mirror operator
on the full plane at $z_2 = \bar{z}_1 = x - iy$.
As $\Phi(z_1)$ approaches the boundary so does its mirror 
and we can use the bulk OPE
\beq
\Phi(z_1) \Phi(z_2) = \sum_i C^{i}_{\Phi\Phi} 
\frac{1}{(z_1- z_2)^{2h_{\Phi} - h_i}}
\frac{1}{(\bar{z}_1- \bar{z}_2)^{2\bar{h}_{\Phi} - \bar{h}_i}}
\psi_{i}\left(\frac{z_1+z_2}{2}\right).
\eeq
on the product $\Phi(z_1)\Phi(\zbar_1)$ (Fig.2). Recalling that the 
correlation function of a field and its mirror consists of the
holomorphic part only this leads  to 
a relation between boundary and bulk operators of the form\cite{cardy2}
\beq
\Phi(z)  =  C^{d}_{\Phi\Phi} (2y)^{\Delta_d-2h_{\Phi}}
(d(x)+c(x)\log y)
+ \sum_i C^{i}_{\Phi\Phi} 
(2y)^{\Delta_i-2h_{\Phi}}
\psi_{i}(x) 
\label{bulkboundary}
\eeq
where we have singled out the logarithmic boundary operators and the sum runs
over the rest. 
The ordinary boundary   operators
 $\psi_i$ are normalized so that they have correlation functions 
\beq \expect{\psi_i(0)\psi_j(x)}=\delta_{ij} x^{-2\Delta_i}\eeq
but we allow the logarithmic operators to have unspecified 
normalizations for reasons that will appear shortly
\bea \expect{d(0)d(x)}&=&(-2\alpha_d\log x +\alpha'_d) x^{-2\Delta_d}\nn\\
 \expect{c(0)d(x)}&=&\alpha_d x^{-2\Delta_d}\nn\\
 \expect{c(0)c(x)}&=&0
\label{boundarylogOPE}
\eea
so we then find that for operators widely separated but close to the boundary
(ie $y\ll x$)
\bea\expect{\Phi(iy)\Phi(x+iy)}&=&(2y)^{-4h_\Phi}\left(\frac{4y^2}{x^2}\right)^{\Delta_d}
\left(C^d_{\Phi\Phi}\right)^2\left(-2\alpha_d\log \frac{x}{y}+\alpha_d'\right)\nn\\
& &+
(2y)^{-4h_\Phi}\sum_i \left(C^i_{\Phi\Phi}\right)^2
\left(\frac{4y^2}{x^2}\right)^{\Delta_i}\label{ope2}\eea
For the operator $\mu(z,\zbar)$ we  can compare this with what 
 the explicit two point function \rf{boundary2ptfn} gives in the same 
regime
\beq \expect{\mu(z_1,\zbar_1)\mu(z_2,\zbar_2)}=(2y)^\half
\left\{  \left(A\log\frac{4y^2}{x^2}+B\right)\sum_{n=0}^\infty a_n\left(\frac{4y^2}{x^2}\right)^n
 +A\sum_{n=1}^\infty b_n \left(\frac
{4y^2}{x^2}\right)^n\right\}\label{MMMM}
\eeq
where the  $a_n$ and $b_n$ are related to the series expansions of the
hypergeometric functions.  This is consistent with
\rf{ope2} with the logarithmic operators duly appearing if $A\ne 0$
together with a stack
of boundary operators of scaling  dimensions which are all
positive integers.  A similar exercise for the $c=0$ model discussed
earlier gives
\beq \expect{\epsilon(z_1,\zbar_1)\epsilon(z_2,\zbar_2)}=(2y)^{-\fivehalves}
\left\{  \left(A\log\frac{4y^2}{x^2}+B\right)\sum_{n=2}^\infty e_n\left(\frac{4y^2}{x^2}\right)^n
 +A\sum_{n=0}^\infty f_n \left(\frac
{4y^2}{x^2}\right)^n\right\}
\eeq
where now $e_n$ and $f_n$ are related to the series expansion 
of the
hypergeometric functions.

\begin{figure}[ht]
\begin{center}
\mbox{\epsfig{file=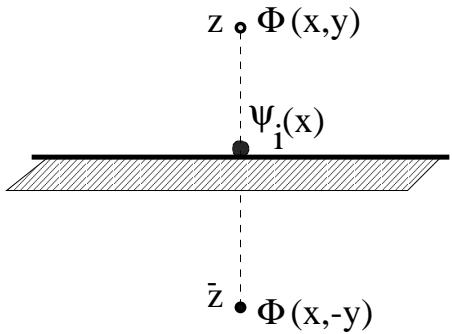,width=6cm}}
\caption { A bulk operator $\Phi(z)$ where $z = x+iy$ 
 induces boundary operators $\psi_i(y)$. In the limit
 $y \rightarrow 0$ this can be seen as an  OPE expansion
  of $\Phi(z)$ and its mirror image $\Phi(\bar{z})$}
\label{Fig.2}
\end{center}
\end{figure}

An obvious question now arises; what
happens to the boundary operators when $A=0$? In this case 
consistency between \rf{MMMM} and \rf{ope2} dictates that 
$\alpha_d$ vanishes but that the coefficient $C^d_{\Phi\Phi}$ does not.
Now the boundary logarithmic operators have correlation functions
\bea \expect{d(0)d(x)}&=&\alpha'_d x^{-2\Delta_d}\nn\\
 \expect{c(0)d(x)}&=&0\nn\\
 \expect{c(0)c(x)}&=&0
\label{boundarylogSC}
\eea
and  the field $c(x)$ has become `sterile' -- it totally decouples
from the rest of the system. 

These results are  very interesting, because they show that, depending
on boundary conditions,  boundary operators may be either logarithmic
or not. This may be related to the fact that D-brane recoil \cite{strings}
(where there are Dirichlet boundary conditions) is described
by logarithmic operators, but there are  no logarithmic operators for ordinary 
open strings (which have Neumann boundary conditions). In this paper we
will not attempt to answer this question in full, but  it seems that the 
fact that  boundary logarithmic operators  may  become non-logarithmic 
under different boundary conditions is important.

Now consider
the limit $ z_1 \rightarrow z_2$, i.e. 
$y >> x$, in the two-point correlation function $\expect{\Phi(z_1)
\Phi(z_2)}$; 
using  the bulk OPE
\beq
\Phi(iy) \Phi(x+iy) =  
\frac{1}{x^{4h_{\Phi} -2 h_C}}
\left(D + C \ln x \right ) + ...
\eeq
 we can relate the expectation values of the logarithmic pair to the
logarithmic  terms in $\expect{\Phi(iy) \Phi(x+iy)}$. Comparing with
the correlation functions (10) and (14) given earlier  immediately tells us that
\beq
\expect{D} =- B \frac{\ln y}{y^{ h_C}}, ~~~~ \expect{C} = B \frac{1}{y^{ h_C}}
\eeq
at least when the scaling dimensions are positive.
Another way of looking at this is directly by considering
 the one-point function
in the presence of a boundary
\bea \expect{D(z)}_{boundary}&=&\expect{D(z)D(\zbar)}\sim\frac{\ln y}{y^{ h_C}}\nn\\
\expect{C(z)}_{boundary}&=&\expect{C(z)C(\zbar)}=0!!
\eea
The calculation of $\expect{C(z)}$ has gone wrong (it violates scale covariance) because of the non-standard
transformation properties of the logarithmic pair. We should consider the LCFT as a limit of an ordinary CFT,   as in \cite{c=0 Quenched Random Magnets and Polymers}, where two ordinary
operators become degenerate and lead to the logarithmic operators; an  operator in the ordinary 
CFT has an image which is itself, but it is a combination of $C$ and $D$ so really we should consider the combination $D+C\log a$ as one operator.

\section{Boundary Conditions and Boundary States in LCFT}
The next step is to investigate the
 connection between boundary conditions, boundary states, and
the $S$ matrix which describes the behaviour of the Virasoro
characters under modular transformations. For ordinary rational
CFTs this  was first elucidated
by Cardy \cite{cardy2} but in the case of the LCFTs his arguments
 are modified by the Jordan cell structure of the Virasoro
generators $L_0$ and ${\overline L}_0$.
At this stage in the development of the subject we do not have a
complete systematic understanding of bulk LCFTs so a corresponding
understanding of boundary conditions and states is impossible. However
we can explore the nature of the differences from ordinary CFTs.

A logarithmic theory occurs when two operators, $O_0(z,\zbar)$ of negative
norm and $O_1(z,\zbar)$ of positive
norm, with weights $(h_0,\bar h_0)$ and $(h_1=h_0+\alpha_D\epsilon^2,
\bar h_1=\bar h_0+\alpha_D\epsilon^2)$ become degenerate as $\epsilon\to 0$.
For simplicity we will assume that $O_{0,1}$ are both primary operators,
 that they and their descendants
are the only  degenerate operators, and that $h=\bar h$. The known 
logarithmic theories are more complicated than this but these assumptions
already lead to significant differences from the non-logarithmic theories.
We can define
\bea D(z,\zbar)&=&\frac{1}{\epsilon}\left(O_0(z,\zbar)+ (1+\frac{\alpha_D'}{2}
\epsilon^2)O_1(z,\zbar)\right)\nn\\
C(z,\zbar)&=&\epsilon\alpha_D O_1(z,\zbar).
\eea
In the limit $\epsilon\to 0$ these operators have the standard 
correlation functions for a logarithmic pair. However, although
$O_0$ and $O_1$ are direct products of holomorphic and anti-holomorphic
sectors, $D$ is not and this affects the Ishibashi states. It is convenient to exploit the ambiguity in the definition of
$D$ to set $\alpha_D'=0$ and to rescale $\epsilon$ and the fields
 so that $\alpha_D=1$. Then we can define the states
\bea \ket{D}&=&\frac{1}{\epsilon}\sum_N \ket{0,N}\otimes\overline{\ket{0,N}}
+\ket{1,N}\otimes\overline{\ket{1,N}}\nn\\
\ket{C}&=&\epsilon\sum_N\ket{1,N}\otimes\overline{\ket{1,N}}\nn\\
\ket{i}&=&\sum_N\ket{i,N}\otimes\overline{\ket{i,N}},\quad i\ge 2\label{Ishibashi}\eea
where the last line is just the standard Ishibashi result \cite{ishibashi} for the
non-logarithmic primary operators $\{O_i,\;i\ge 2\}$.
We can compute the action of $L_0$ on these states. There is one subtlety which
is that since we are going to vary $\epsilon$ 
 we are not entitled
to assume that $\ket{0,N}$ and $\ket{1,N}$ are normalized to a constant;
rather  we 
expect that they have a normalization which is a polynomial in $h$, or
equivalently in $\epsilon^2$ , which we denote $P^{(0,1)}_N(\epsilon^2)$. 
Note that the $P^{(0,1)}_N(\epsilon^2)$ have the property that if
they are non zero at $\epsilon=0$ for a particular descendant $N$ then
 $P^{(0)}_N(0)=- P^{(1)}_N(0)$. We find that
\bea \bra{D} q^{L_{0} - \frac{c}{24}}\ket{D}&=&
\lim_{\epsilon\to 0}\epsilon^{-2}\sum_N P^{(0)}_N(\epsilon^2)
q^{h_{0}+N - \frac{c}{24}}+P^{(1)}_N(\epsilon^2)
q^{h_{0}+\epsilon^2+N - \frac{c}{24}}\nn\\
&=&\chi_0(q)\log q+\chi_1(q)\nn\\
\bra{D} q^{L_{0} - \frac{c}{24}}\ket{C}&=&
\lim_{\epsilon\to 0}\sum_NP^{(1)}_N(\epsilon^2)
q^{h_{0}+\epsilon^2+N - \frac{c}{24}}\nn\\
&=&\chi_0(q)\nn\\
\bra{C} q^{L_{0} - \frac{c}{24}}\ket{C}&=&0\label{Telements}\eea
where
\bea \chi_0(q)&=&\sum_N P^{(1)}_N(0)q^{h_{0}+N - \frac{c}{24}}\nn\\
\chi_1(q)&=&\lim_{\epsilon\to 0}\sum_N\epsilon^{-2}\left(P^{(0)}_N(\epsilon^2)
+P^{(1)}_N(\epsilon^2)\right)q^{h_{0}+N - \frac{c}{24}}.\label{logchars}\eea
Note that on account of the properties of the $P^{(0,1)}_N(\epsilon^2)$ the limit in \rf{logchars} exists.
Furthermore the character $\chi_1(q)$ has by definition no contribution
at level $N=0$ so it appears to belong to a representation with conformal
weight one higher than does $\chi_0(q)$. We still have the same number
of independent character functions; the only exception to this would be
if it happened that $\chi_1(q)=0$ but this does not occur in the only case
where the characters are known explicitly (see below).

Now, following Cardy consider the region formed by 
identifying the edges $\Re z=0$ and $\Re z=2\pi\Im\tau$ (where $\tau$ is
taken to be imaginary) of the rectangular
region $0< \Re z< 2\pi\Im\tau$, $0<\Im z<\pi$. This can be viewed either as
an annulus in which states propagate in the $\Re z$ direction or
as a cylinder in which states propagate in the $\Im z$ direction (Fig.3).
This construction is familiar in string theory where the same process can
be described either as the propagation of open strings (annulus) or 
of closed strings (cylinder). Imposing
boundary conditions labelled $\alpha$ and $\beta$ on the annulus
configuration then corresponds to evolution on the cylinder configuration
with initial state $\ket\beta$ and final state $\ket\alpha$.
Under the conformal
transformation $\xi=\exp(-iz/\Im\tau)$ the infinite  cylinder
of which our cylinder is a segment becomes the whole plane and  therefore
the transfer matrix in the
$\Im z$ direction is given by the Virasoro generators on the plane. 
Writing
\bea \ket{\alpha}&=&a\ket{D}+a'\ket{C}+\sum_{i\ge 2}\alpha_i\ket{i}\nn\\
\ket{\beta}&=&b\ket{D}+b'\ket{C}+\sum_{i\ge 2}\beta_i\ket{i}\eea
 and setting
$ \tilde q=\exp(-2\pi i/\tau)$ we find that 
the matrix elements of the transfer matrix take the form
\bea Z_{\alpha \beta}&=&\bra{\alpha}\tilde q^{\frac{1}{2}\left(L_{0}^{c}+ \bar{L}_{0}^{c}\right) - \frac{c}{24}}\ket{\beta}\nn\\
&=& ab\left(\chi_0(\tilde q)\log\tilde q+\chi_1(\tilde q)\right)
\nn\\
&&+\left(ab'+ba'\right)\chi_0(\tilde q)+\sum_{i\ge 2}\alpha_i\beta_i\chi_i(\tilde q).
\label{cylinderZ}\eea

\begin{figure}[ht]
\begin{center}
\mbox{\epsfig{file=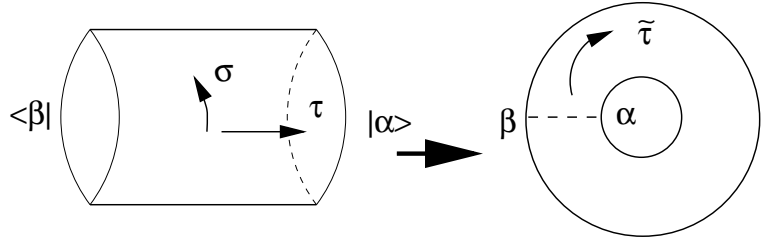,width=10cm}}
\caption {The relationship between cylinder and annulus.}
\label{Fig.3}

\end{center}
\end{figure}

Now we calculate the partition function by considering the transfer 
matrix in the $\Re z$ direction i.e. round the annulus. We identify  the
states at $\Re z=0$ and $\Re z=\pi$ and then sum over all of them
to obtain
\bea
Z_{\alpha \beta} 
&=&Tr_{\alpha \beta} {q}^{L_{0}^{o} - \frac{c}{24}}\nn\\
&=&\sum_i n^{i}_{\alpha\beta}
\chi_i({q})\label{annulusZ}
\eea
where $q=\exp(i2\pi\tau)$ and $n^{i}_{\alpha\beta}$ is the number of times the representation $i$
 occurs in the spectrum of the boundary theory
with  two boundaries and boundary conditions $\alpha$ and $\beta$.
Note that $\log q$ does not appear in \rf{annulusZ}.

Now $q$ is related to $\tilde q$ by  $\tau \to - 1/\tau$ and so we need to know the behaviour of the characters under  
a modular transformation.   From the fact that the partition function calculated on the cylinder \rf{cylinderZ} and on the annulus \rf{annulusZ} must be equal we see that the characters should transform as
\beq \chi_i(q)=\sum_{j} \left(S_{i}^j+\frac{\log\tilde q}{2\pi}Q_{i}^j\right)
\chi_j(\tilde q).\label{characterS}\eeq
Consistency then requires
\beq S^2+Q^2=1, QS=SQ=0\eeq
which in turn implies that if $Q\ne 0$ then both $\det S=0$ and $\det Q=0$.
Equating \rf{annulusZ} and \rf{cylinderZ} 
we obtain the relationships
\bea ab&=&\frac{1}{2\pi} n^{i}_{\alpha\beta}Q_i^0= n^{i}_{\alpha\beta}S_i^1\nn\\
a'b+ab'&=&n^{i}_{\alpha\beta}S_i^0\nn\\
\alpha_j\beta_j&=&n^{i}_{\alpha\beta}S_i^j,\quad j\ge 2\label{Cgens}\eea

The only case where the characters are known explicitly is the $c=-2$
model \cite{fusionrules1,fusionrules2,fusionrules3,c=-2}; these characters do indeed transform according to \rf{characterS}. The construction we have outlined works if we identify our characters
in the following way
\bea \chi_0(q)&=&\frac{\partial\Theta_{1,2}(q)}{\eta(q)}\nn\\
\chi_1(q)&=&\frac{1}{\eta(q)}\left(\Theta_{1,2}(q)-\partial\Theta_{1,2}(q)
\right)\eea
and $\chi_i$, $i=2,3$ are the characters for the normal fields
with conformal weights $h=-1/8$ and $h=3/8$ respectively.
In terms of the space of states constructed in \cite{fusionrules2}
$\half\chi_1(q)$ is the character for $\nu_1$ and
$\chi_0(q)+\half\chi_1(q)$ is the character for $\nu_0$.
Then $S$ and $Q$ are given by
\beq S=\left(\begin{array}{cccc}0&0&0&0\\
0&0&\half&-\half\\
1&1&\half&\half\\
-1&-1&\half&\half\\\end{array}\right),
\qquad Q=\left(\begin{array}{cccc}1&0&0&0\\
-1&0&0&0\\
0&0&0&0\\
0&0&0&0\\\end{array}\right).\label{SQdefs}\eeq
There are solutions to the equations \rf{Cgens} in this case but they do
not take the simple form found by Cardy for the unitary minimal models. 
In particular there is no boundary state $\ket{\tilde k}$ for which 
just one highest weight representation contributes to the annulus
amplitude ie for which $n^{i}_{\tilde k\tilde k}=\delta^i_k$
for some $k$.  The presence of the factor of $2\pi$ in (41) and the
fact that $S$ and $Q$ satisfy (40) implies that $ab=0$; furthermore 
the first two columns of $S$ are identical so $a'b+ab'=0$ too. 
(We note that in this construction the presence of the $2\pi$ factor
appears unavoidable.)
The $n^{i}_{\alpha\beta}$ must satisfy
$n^{0}_{\alpha\beta}=n^{1}_{\alpha\beta}$ and
$n^{2}_{\alpha\beta}=n^{3}_{\alpha\beta}$. If we try to impose the same
boundary condition on each boundary ie $\ket{\alpha}=\ket{\beta}$ then
$a=0$ and $a'$ (which is the coefficient of a zero-norm state)
is undetermined; in addition
\bea \alpha_2^2&=&n^{2}_{\alpha\alpha}+\half n^{1}_{\alpha\alpha} \nn\\
\alpha_3^2&=&n^{2}_{\alpha\alpha}-\half n^{1}_{\alpha\alpha}. \eea
There are no non-trivial solutions when $n^{2}_{\alpha\alpha}=0$ but 
if $n^{2}_{\alpha\alpha}=1$ then $n^{1}_{\alpha\alpha}=0,1,2$ are allowed
and we get the states
\bea \ket{\tilde 1}&=&a'\ket{C}+\ket{2}+\ket{3}\nn\\
\ket{\tilde 2}&=&a''\ket{C}+\sqrt{\frac{3}{2}}\ket{2}+\sqrt{\frac{1}{2}}\ket{3}\nn\\
\ket{\tilde 3}&=&a'''\ket{C}+\sqrt{2}\ket{2}.\eea
These are linearly independent because of the presence of the 
zero-norm state but not orthogonal.

\section{Conclusions}
In this paper we have discussed how the properties of boundary LCFTs
depend very delicately on the boundary conditions and are quite different
from those of the same LCFT without boundaries. Operators which in the
pure bulk theory do not have logarithmic two-point functions 
(but do have logarithmic four-point functions) acquire 
logarithmic two-point functions in the presence of a boundary; the logarithms
show up either in the bulk, or close to the boundary, or both depending
upon the boundary conditions.  Whether or not there are boundary logarithmic
operators also depends on the boundary conditions. We have discussed  how the 
Cardy conditions relating boundary states and bulk quantities are modified
in LCFTs.
\vskip1cm
We acknowledge the support of PPARC grant PPA/G/O/1998/00567
and stimulating discussions with John Cardy, Jean-Sebastien Caux,
and Nick Mavromatos, and the comments of Victor Gurarie and the referee.

\end{document}